\documentclass[]{l4dc2026}

\title[Learning Nonlinear Nonholonomic Dynamics]{Structure-Preserving Learning for Nonlinear Nonholonomic Dynamics}
\usepackage{times}
\usepackage{booktabs}

\usepackage{mathtools}

\newlength{\myeqskip}  
\AtBeginDocument{%
    \setlength\abovedisplayskip{\myeqskip}%
    \setlength\belowdisplayskip{\myeqskip}%
    \setlength\abovedisplayshortskip{\myeqskip-\baselineskip}%
    \setlength\belowdisplayshortskip{\myeqskip}}

\author{%
 \Name{Thomas Beckers} \Email{thomas.beckers@uni-ulm.de}\\
 \addr Institute of Measurement, Control and Microtechnology, Ulm University, Ulm, Germany
 \AND
 \Name{Anthony Bloch} \Email{abloch@umich.edu}\\
 \addr Department of Mathematics, University of Michigan, Ann Arbor, MI 48109, USA
 \AND
 \Name{Leonardo Colombo} \Email{leonardo.colombo@csic.es}\\
 \addr Centre for Automation and Robotics (CAR), Spanish National Research Council (CSIC), Madrid, Spain
}
\begin{document}

\maketitle

\begin{abstract}
Learning the dynamics of mechanical systems subject to nonlinear nonholonomic constraints is challenging, as data-driven models may violate the underlying physical constraints, leading to inconsistent predictions. To address this problem, we develop a constraint-preserving learning framework based on the Lagrange–d’Alembert–Chetaev equations and tangency to the constraint submanifold, which together define an affine space of admissible accelerations. Using the kinetic-energy metric and the Moore–Penrose pseudoinverse, we construct a metric projector and the corresponding Chetaev correction. Embedding this geometric structure directly into a Gaussian process prior ensures that both prior and posterior dynamics satisfy the acceleration constraints exactly. We characterize the resulting reproducing kernel Hilbert space (RKHS), derive an adapted-coordinate representation, and establish nonexpansiveness and consistency transfer. Numerical experiments on a three-dimensional directional flocking system with up to $N=50$ agents demonstrate exact constraint preservation and robustness to constraint-incompatible observations.
\end{abstract}

\section{Introduction}
\label{sec:introduction}

Learning dynamical models from data has become increasingly important in
robotics and control, particularly when accurate first-principles models are
unavailable or affected by uncertainty~\citep{Brunton_Kutz_2019}.  In this context, Gaussian processes (GPs) have shown promising results in learning dynamical models from data, offering a flexible probabilistic framework that combines nonparametric regression with uncertainty quantification~\citep{RasmussenWilliams2006}. However, unconstrained learning methods generally do not preserve the underlying geometric structure of mechanical systems, potentially limiting their generalization capabilities and reliability. This has
motivated structure-preserving approaches, e.g., based on Hamiltonian and Lagrangian
models \citep{GreydanusEtAl2019,CranmerEtAl2020} and physics-informed GPs for mechanical and
port-Hamiltonian systems
\citep{pmlr-v120-geist20a, BeckersEtAl2022PortHamiltonian}. 

An especially interesting class of physical systems is that of nonholonomic mechanics, in which geometric constraints restrict the admissible motions and fundamentally shape the system dynamics.  Representative examples include wheeled mobile robots, car-like vehicles, rolling disks, and robotic systems subject to non-slip contact conditions~\citep{bloch2004nonholonomic}.  Learning such systems without accounting for this
structure may produce physically inconsistent predictions. Recent
approaches to overcome this issue include constraint discovery \citep{WangBloch2024} and Lagrangian 
networks \citep{DiazEtAl2025}. In particular,
\citep{BeckersBlochColombo2026} embed a linear nonholonomic distribution
directly into a matrix-valued GP kernel. More generally, constraint-preserving GP
kernels for known linear operator constraints were developed in
\cite{jidling2017linearly}. 

However, \emph{nonlinear} nonholonomic constraints are geometrically different: they define a constraint submanifold of the tangent bundle together with a velocity-dependent space of admissible directions
\citep{DeLeonMarreroDiego1997,StratoglouEtAl2025TAC}. Typical examples include constant-kinetic-energy constraints
(thermostats) and nonlinear constraints imposing parallelism between
velocities \citep{BlochRojo2016}. Unfortunately, the
projection construction used for linear nonholonomic constraints in
\citep{BeckersBlochColombo2026} does not apply directly. The nonlinear case is
not simply obtained by replacing the linear projector by a state-dependent as second-order tangency yields an affine constraint $Aa+b=0$ for admissible accelerations $a$. Moreover, natural
representations of nonlinear nonholonomic constraints may involve
constraint functions that are not independent, leading to singular
multiplier equations, e.g., by cross-product descriptions of
parallel-velocity constraints in flocking systems. 

\textbf{Main contributions.}
We develop a structure-preserving GP framework for learning nonlinear 
nonholonomic dynamics whose constraint functions need not be independent.
The Chetaev equations and Moore-Penrose pseudoinverse yield an affine
decomposition of admissible accelerations that is independent of the
chosen constraint representation and can be embedded directly into a
matrix-valued GP prior. We prove positive semidefiniteness, exact
constraint preservation by prior and posterior predictions, characterize
the associated RKHS and adapted-coordinate representation, and establish
nonexpansiveness and consistency transfer in the kinetic-energy metric.
We also distinguish constrained-prior conditioning from posterior projection, showing that the former exactly excludes constraint-incompatible observation components that may persist in the latter. We extend the framework to 3D directional flocking with nonlinear cross-product constraints and establish a common metric projector for connected velocity-alignment graphs. Experiments with up to 50 agents show that the proposed nonholonomic GP outperforms standard and projected GPs.

The remainder of the paper is organized as follows. Section~\ref{subsec:nonlinear_mechanics} reviews nonlinear nonholonomic mechanics.   Section~\ref{sec:learning} develops the structure-preserving
kernel and its theoretical properties. Section~\ref{sec:flocking} presents the
three-dimensional flocking application, followed by concluding remarks.


\section{Nonlinear Nonholonomic Mechanics}
\label{subsec:nonlinear_mechanics}

Let $Q$ be an $n$-dimensional configuration manifold endowed with a
Riemannian metric $\mathbb G$, and consider the mechanical Lagrangian
$L(v_q)=\frac12\mathbb G_q(v_q,v_q)-V(q)$, $v_q\in T_qQ$.
In local coordinates, $\mathbb G$ is represented by a symmetric
positive-definite matrix $M(q)$, and we write $x=(q,v)\in TQ$. Throughout, $D_q$ and $D_v$ denote
partial derivatives with respect to the generalized position $q$ and velocity $v$, respectively. The system is subject to $m$ nonlinear velocity constraints
$\Phi(q,v)=0$, with $\Phi:TQ\to\mathbb{R}^m$. We restrict our attention
to a smooth submanifold $\mathcal C\subset\Phi^{-1}(0)$ satisfying
$T_x\mathcal C=\ker D\Phi(x)$ for every $x\in\mathcal C$, see \cite{DeLeonMarreroDiego1997,CendraEtAl2004}, while the defining
constraint functions need not be independent. We assume that $A(q,v):=D_v\Phi(q,v)$ has constant rank $r\le m$ on $C$
and that $D_q\Phi(q,v)v\in\operatorname{Im}A(q,v)$.

Unlike the linear nonholonomic case,
the admissible velocities at fixed position $q$ generally do not form a vector
space. Instead, each $x=(q,v)\in\mathcal C$ determines the
velocity-dependent subspace
$\mathcal S_x:=\ker A(q,v)\subset T_qQ$,
of dimension $n-r$,
which is the fiberwise
linearization of the nonlinear constraints, see 
\cite{StratoglouEtAl2025TAC}.

Then, the dynamics is described by the Lagrange-d'Alembert-Chetaev
equations
\begin{equation}
    \frac{d}{dt}\frac{\partial L}{\partial v}
    -\frac{\partial L}{\partial q}
    =A(q,v)^\top\lambda,
    \qquad
    \Phi(q,v)=0,
    \label{eq:chetaev}
\end{equation}
where $\lambda\in\mathbb{R}^m$ are the Lagrange multipliers
(\cite{CendraEtAl2004}, \cite{deLeon2012Historical}). Via the kinetic-energy metric, the Chetaev reaction covector
$A^\top\lambda$ corresponds locally to the acceleration
$M^{-1}A^\top\lambda\in S_x^{\perp_M}$, where  $\langle z,w\rangle_M:=z^\top M(q)w$, 
$S_x^{\perp_M}:=\{z\in T_qQ:\langle z,w\rangle_M=0
\ \forall\,w\in S_x\}$.

Let $g(q,v)$ denote the unconstrained acceleration, so that
$\dot v=g+M^{-1}A^\top\lambda$. Tangency to $\mathcal C$ requires
$D_q\Phi\,v+D_v\Phi\,\dot v=0$. Setting
$b:=D_q\Phi\,v$ and $W:=AM^{-1}A^\top$ gives
$W\lambda=-(Ag+b)$. Since $M^{-1}$ is positive definite, $\ker W=\ker A^\top$ and hence
$\operatorname{Im}W=\operatorname{Im}A$; therefore the condition
$b\in\operatorname{Im}A$ guarantees solvability. Every solution is of the
form $\lambda=-W^\dagger(Ag+b)+\zeta$, with
$\zeta\in\ker W=\ker A^\top$, so $A^\top\lambda$ is independent of
$\zeta$. Substitution yields
\begin{equation}
a_\Phi(x)=c_\Phi(x)+\Pi_\Phi(x)g(x),\qquad
\Pi_\Phi=I-M^{-1}A^\top W^\dagger A,\qquad
c_\Phi=-M^{-1}A^\top W^\dagger b,
\label{eq:chetaev_decomposition}
\end{equation}
where $a_\Phi(x)$ denotes the acceleration process. By the constant-rank assumption, $W^\dagger$, and hence
$\Pi_\Phi$ and $c_\Phi$, vary smoothly on $\mathcal C$. Here $\Pi_\Phi$ is the $M(q)$-orthogonal projector onto $\ker A(x)$,
while $Ac_\Phi=-b$ and
$c_\Phi\in\operatorname{Im}(M^{-1}A^\top)
=(\ker A)^{\perp_M}$.
Hence $c_\Phi$ is the minimum-$M(q)$-norm solution of $Aa+b=0$, and
the admissible accelerations form
$\mathcal A_x=c_\Phi(x)+\ker A(x)$.

If $\Phi$ depends only on $v$, then $D_q\Phi=0$ and therefore
$c_\Phi=0$. The admissible acceleration space reduces to
$\ker D_v\Phi(v)$, although $\Pi_\Phi(v)$ remains velocity dependent
when $\Phi$ is nonlinear. This is the setting considered in the
flocking application of Section~\ref{sec:flocking}.


\section{Structure-Preserving Learning of Nonlinear Nonholonomic Dynamics}
\label{sec:learning}
Next, we propose our learning approach for Lagrange-d’Alembert-Chetaev systems~\eqref{eq:chetaev}. The general idea is to use the decomposition~\eqref{eq:chetaev_decomposition}
to equip an ambient GP with a prior whose mean, covariance, and
associated RKHS respect the nonlinear nonholonomic structure by construction. For that, let
$g:\mathcal C\to\mathbb R^n$ on a fixed coordinate domain $\mathcal C$ and consider
$g\sim\mathcal{GP}(m_0,K_0)$, where
$m_0:\mathcal C\to\mathbb R^n$ is the mean function and
$K_0:\mathcal C\times\mathcal C\to\mathbb R^{n\times n}$ is a
matrix-valued covariance kernel. 
A simple choice used throughout the paper is
$K_0(x,x')=k(x,x')I_n$, where $k$ is a scalar positive-definite
kernel. Associated with $K_0$ is a vector-valued RKHS $\mathcal H_{K_0}$
\citep{RasmussenWilliams2006}.

\subsection{Nonlinear Nonholonomic Kernel}
\label{subsec:nh_kernel}

Let $g\!\sim\!\mathcal{GP}(m_0,K_0)$ be an ambient GP as introduced before. We define the 
acceleration process by
\begin{equation}
    a_\Phi(x):=c_\Phi(x)+\Pi_\Phi(x)g(x).
    \label{eq:constrained_process}
\end{equation}
\begin{proposition}
\label{prop:gp_pushforward}
The process $a_\Phi$ is a GP, $a_\Phi\sim\mathcal{GP}(m_\Phi,K_\Phi)$, with
$m_\Phi(x)=c_\Phi(x)+\Pi_\Phi(x)m_0(x)$ and
$K_\Phi(x,x')\!\!=\!\!\Pi_\Phi(x)K_0(x,x')\Pi_\Phi(x')^\top\!\!$, where $K_\Phi$ is a positive-semidefinite matrix-valued kernel.
\end{proposition}
\begin{proof}
For $X=(x_1,\ldots,x_N)\subset\mathcal C$, let
$G_X:=\operatorname{col}_i g(x_i)$,
$C_X:=\operatorname{col}_i c_\Phi(x_i)$, and
$P_X:=\operatorname{diag}_i\Pi_\Phi(x_i)$.
By~\eqref{eq:constrained_process},
$\operatorname{col}_i a_\Phi(x_i)=C_X+P_XG_X$.
Since $G_X$ is jointly Gaussian, this vector is Gaussian with mean
$C_X+P_Xm_0(X)$ and covariance $P_XK_0(X,X)P_X^\top$,
whose $(i,j)$ block is $K_\Phi(x_i,x_j)$ and which is positive
semidefinite because $K_0(X,X)\succeq0$.
\end{proof}
For the ambient kernel
$K_0(x,x')=k(x,x')I_n$, the structured kernel then becomes
$K_\Phi(x,x')
=\Pi_\Phi(x)k(x,x')\Pi_\Phi(x')^\top$. Since $\Pi_\Phi$ is the unique $M(q)$-orthogonal projector onto
$\ker A$ and $c_\Phi$ is the unique minimum-$M(q)$-norm element of
$\mathcal A_x$, the construction is invariant under different defining maps of the
same admissible acceleration space.
Consequently, $m_\Phi$ and $K_\Phi$ depend only on
$\mathcal A_x$, and not on the particular choice of constraint
equations. Next, we show that this prior almost surely guarantees constraint satisfaction.
\begin{theorem}
\label{thm:constraint_preservation}
For every $x,x'\in\mathcal C$, $A(x)m_\Phi(x)+b(x)=0$ and $A(x)K_\Phi(x,x')=0$. Hence the prior satisfies
$A(x)a_\Phi(x)+b(x)=0$ almost surely.
Moreover, for observations
$y_i=a_\Phi(x_i)+\varepsilon_i$ with independent
$\varepsilon\sim\mathcal N(0,\Sigma)$, $\Sigma\succ0$, the posterior
mean $\hat a_\Phi$ and covariance $K_{\Phi,N}$ satisfy $A(x)\hat a_\Phi(x)+b(x)=0$ and $A(x)K_{\Phi,N}(x,x')=0$. Hence posterior realizations also satisfy
$A(x)a_\Phi(x)+b(x)=0$ almost surely.
\end{theorem}
\begin{proof}
Since $A\Pi_\Phi=0$ and $Ac_\Phi=-b$, the prior identities follow
immediately from
$m_\Phi=c_\Phi+\Pi_\Phi m_0$,
$K_\Phi(x,x')=\Pi_\Phi(x)K_0(x,x')\Pi_\Phi(x')^\top$, and
$a_\Phi=c_\Phi+\Pi_\Phi g$.
Let $Q_X:=K_\Phi(X,X)+\Sigma$.
The posterior mean and covariance are
$\hat a_\Phi(x)=m_\Phi(x)+K_\Phi(x,X)Q_X^{-1}(Y-m_\Phi(X))$ and
$K_{\Phi,N}(x,x')=
K_\Phi(x,x')-K_\Phi(x,X)Q_X^{-1}K_\Phi(X,x')$.
Since $A(x)K_\Phi(x,X)=0$, left multiplication by $A(x)$ gives the
posterior identities.
\end{proof}
The theorem also has a direct dynamical interpretation. Defining the
second-order vector field
$F_\Phi(q,v):=(v,\widehat a_\Phi(q,v))$, one obtains
$D\Phi(x)F_\Phi(x)
    =
    D_q\Phi(q,v)v+D_v\Phi(q,v)\widehat a_\Phi(q,v)
    =
    b(x)+A(x)\widehat a_\Phi(x)
    =0$. Therefore $F_\Phi(x)\in T_x\mathcal C$ for every $x\in\mathcal C$.

\paragraph{Conditioning versus post-projection.}
The construction above is generally different from conditioning an ambient GP
and projecting its posterior afterwards as explained, e.g., in~\cite{geist2021structured}. To make this distinction explicit,
consider the centered case $m_0\equiv 0$ and $c_\Phi\equiv 0$ and let
$\mathbf P_X:=\operatorname{diag}(\Pi_\Phi(x_1),\ldots,\Pi_\Phi(x_N))$.
Then the posterior mean induced by the constrained prior is $\widehat a_{\Phi,N}(x)
=
\Pi_\Phi(x)K_0(x,X)\mathbf P_X^\top
\Big(
\mathbf P_X K_0(X,X)\mathbf P_X^\top+\Sigma
\Big)^{-1}Y$. By contrast, conditioning the ambient GP first gives $\widehat g_N(x)
=
K_0(x,X)
\big(K_0(X,X)+\Sigma\big)^{-1}Y$, and $\widehat a_{\mathrm{PP,N}}(x):=\Pi_\Phi(x)\widehat g_N(x)$.
Similarly, one may project the  posterior covariance according to $K_{\mathrm{ PP,N}}(x,x')=\Pi_\Phi(x)K_{0,N}(x,x')\Pi_\Phi(x')^\top$. Thus the post-projected posterior also satisfies the homogeneous constraint
pointwise, but it does not in general coincide with the posterior obtained by
conditioning the constrained prior as we proposed. The two procedures differ because the
constraint geometry enters \emph{before} conditioning in the present construction and \emph{after} conditioning in the post-projected model.

\paragraph{Rejection of constraint-incompatible observation components.}
The distinction above is particularly relevant in the presence of measurement
noise. For fixed kernel and noise hyperparameters, assume for simplicity that
$\Sigma=\sigma^2 I$ and consider a perturbation $\eta$ of the stacked observations satisfying
$\mathbf P_X^\top\eta=0$. Then the posterior mean obtained from the constrained
prior is unchanged:
$\widehat a_{\Phi,N}[Y+\eta](x)
=
\widehat a_{\Phi,N}[Y](x)$. Indeed, $K_\Phi(X,X)\eta=0$ and hence
$(K_\Phi(X,X)+\sigma^2I)^{-1}\eta=\sigma^{-2}\eta$, while
$K_\Phi(x,X)\eta=0$. By contrast, the associated change in the
post-projected estimator is
\begin{align}
\widehat a_{\mathrm{ PP,N}}[Y+\eta](x)-\widehat a_{\mathrm{ PP,N}}[Y](x)
=
\Pi_\Phi(x)K_0(x,X)
\bigl(K_0(X,X)+\sigma^2I\bigr)^{-1}\eta,
\end{align}
which need not vanish. Thus the constrained prior removes observation
components incompatible with admissible directions at training states
before they can be propagated by the regression model.

When $\Pi_\Phi$ is Euclidean orthogonal, in particular for $M=I$, the condition
$\Pi_\Phi(x_i)^\top\eta_i=0$ means that $\eta_i$ is normal to the admissible
acceleration directions at $x_i$; the analogous interpretation holds in
kinetic-energy coordinates for a general mass metric.

\subsection{RKHS Characterization and Adapted Coordinates}
\label{subsec:rkhs}
In the following, we analyze the reproducing kernel Hilbert space (RKHS) associated with the nonlinear nonholonomic kernel introduced in Proposition~\ref{prop:gp_pushforward}. 
Let $\mathcal H_{K_0}$ and $\mathcal H_\Phi$ denote the RKHS
associated with $K_0$ and $K_\Phi$, respectively, and define the
linear operator $(Tg)(x):=\Pi_\Phi(x)g(x)$.
\begin{theorem}
\label{thm:rkhs}
The RKHS associated with $K_\Phi$ is
$\mathcal H_\Phi
    =
    \{
        Tg:\,g\in\mathcal H_{K_0}
    \}
    =
    \{
        \Pi_\Phi(\cdot)g(\cdot):
        g\in\mathcal H_{K_0}
    \}$, equipped with the quotient norm
$ \|h\|_{\mathcal H_\Phi}
    =
    \inf
    \left\{
        \|g\|_{\mathcal H_{K_0}}:
        Tg=h
    \right\}$. In particular,
$A(x)h(x)=0$ for every
$h\in\mathcal H_\Phi$ and $x\in\mathcal C$.
\end{theorem}
\begin{proof}
Let
$\mathcal N:=\ker T
=\{g\in\mathcal H_{K_0}:
\Pi_\Phi(x)g(x)=0,\ \forall x\in\mathcal C\}$.
Since point evaluation is continuous in an RKHS,
each map $g\mapsto\Pi_\Phi(x)g(x)$ is continuous, and therefore
$\mathcal N$ is closed. Hence
$\mathcal H_{K_0}=\mathcal N\oplus\mathcal N^\perp$. Equipping $T(\mathcal H_{K_0})$ with the quotient norm identifies it
isometrically with $\mathcal N^\perp$ through
$T|_{\mathcal N^\perp}$; in particular,
$T(\mathcal H_{K_0})$ is complete.

For $x\in\mathcal C$ and $c\in\mathbb R^n$, define
$h_{x,c}
    :=
    K_0(\cdot,x)\Pi_\Phi(x)^\top c
    \in\mathcal H_{K_0}$. For any $n\in\mathcal N$, the reproducing property gives $\langle n,h_{x,c}\rangle_{\mathcal H_{K_0}}
    =
    c^\top\Pi_\Phi(x)n(x)
    =0$, so $h_{x,c}\in\mathcal N^\perp$.

Moreover, $(Th_{x,c})(x')
    =
    \Pi_\Phi(x')
    K_0(x',x)
    \Pi_\Phi(x)^\top c=
    K_\Phi(x',x)c$. Thus the kernel sections of $K_\Phi$ are precisely the images under
$T$ of the functions $h_{x,c}$.

We show that the span of the $h_{x,c}$ is dense in
$\mathcal N^\perp$. Suppose
$g\in\mathcal N^\perp$ is orthogonal to all $h_{x,c}$. For every
$x$ and $c$, $0
    =
    \langle g,h_{x,c}\rangle_{\mathcal H_{K_0}}
    =
    c^\top\Pi_\Phi(x)g(x)$. Since this holds for all $c\in\mathbb R^n$,
$\Pi_\Phi(x)g(x)=0$ for all $x$, so $g\in\mathcal N$.
Because also $g\in\mathcal N^\perp$, we conclude $g=0$.
Hence the span is dense in $\mathcal N^\perp$.

Hence, the span of the kernel sections
$K_\Phi(\cdot,x)c=T h_{x,c}$ is dense in
$T(\mathcal H_{K_0})$ with the quotient norm. So,
$T(\mathcal H_{K_0})$ is precisely the RKHS associated with $K_\Phi$.
Equivalently, two elements $g_1,g_2\in\mathcal H_{K_0}$ represent the
same element of $\mathcal H_\Phi$ if and only if
$g_1-g_2\in\mathcal N$, which gives exactly the norm $\|h\|_{\mathcal H_\Phi}$. Finally,
$A(x)h(x)=A(x)\Pi_\Phi(x)g(x)=0$ follows from
$A(x)\Pi_\Phi(x)=0$.
\end{proof}
Theorem~\ref{thm:rkhs} characterizes the centered, constraint-compatible component of
the model: when $m_0\equiv0$, posterior means belong to the affine space
$c_\Phi+\mathcal H_\Phi$, while $\mathcal H_\Phi$ itself depends only on
$\ker A$ and is therefore independent of the chosen constraint
representation. To obtain a local reduced-coordinate description, let
$\mathcal U\subset\mathcal C$ be an open set on which $\ker A(x)$ admits
a smooth frame $B(x)\in\mathbb R^{n\times(n-r)}$, with full column rank
and $\operatorname{Im}B(x)=\ker A(x)$. Next, we establish a unique reduced-dimensional representation of the constrained acceleration process and a corresponding factorization of its kernel. This factorization provides the basis for characterizing the RKHS associated with the nonlinear nonholonomic kernel in terms of the RKHS of the reduced GP.
\begin{corollary}
\label{cor:adapted}
On $\mathcal U$, let
$L_B(x):=
\bigl(B(x)^\top M(q)B(x)\bigr)^{-1}B(x)^\top M(q)$. Then $\Pi_\Phi(x)=B(x)L_B(x)$, and every realization of the constrained acceleration process admits the
unique representation
\begin{equation}
a_\Phi(x)=c_\Phi(x)+B(x)\nu(x),
\qquad
\nu(x)=L_B(x)g(x).
\label{eq:adapted}
\end{equation}
Moreover, $\nu$ is an $(n-r)$-dimensional GP with $m_\nu(x)\!=\!L_B(x)m_0(x)$ and kernel matrix $K_\nu(x,x')
\!=\!
L_B(x)K_0(x,x')L_B(x')^\top$. The constrained kernel factorizes as
$K_\Phi(x,x')
\!=\!
B(x)K_\nu(x,x')B(x')^\top$.
\end{corollary}
\begin{proof}
Since $B(x)$ has full column rank and $M(q)$ is positive definite,
$B(x)^\top M(q)B(x)$ is positive definite. Hence $L_B(x)$ is well defined
and $L_B(x)B(x)=I$. The map
$B(x)L_B(x)$ is the $M(q)$-orthogonal projector onto
$\operatorname{Im}B(x)=\ker A(x)$, and therefore equals $\Pi_\Phi(x)$.
Thus $a_\Phi-c_\Phi=\Pi_\Phi g=B L_B g$,
which gives~\eqref{eq:adapted}; uniqueness follows from the injectivity of
$B(x)$. Since $\nu=L_Bg$ is a linear transformation of the ambient GP, its
mean and covariance are $m_\nu$ and $K_\nu$ above. The factorization of
$K_\Phi$ follows from $\Pi_\Phi=B L_B$.
\end{proof}
Although the reduced coordinates depend on the chosen local frame, the
resulting constrained process does not. Indeed, if
$\widetilde B=BR$, with $R(x)$ invertible, then
$\widetilde\nu=R^{-1}\nu$, while $a_\Phi$ and $K_\Phi$ remain unchanged.
For $M=I$, choosing an orthonormal frame $B(x)=U(x)$ gives
$\Pi_\Phi(x)=U(x)U(x)^\top$ and, for
$K_0(x,x')=k(x,x')I_n$, it leads to
$K_\nu(x,x')=k(x,x')U(x)^\top U(x')$.
This is the form used in the flocking application of Section~\ref{sec:flocking}.
\subsection{Nonexpansiveness and Consistency Transfer}
\label{subsec:consistency}

We next show that the Chetaev projection does not amplify learning
errors when measured in the kinetic-energy metric. Define
$\|z\|_{M(q)}^2:=z^\top M(q)z$.
\begin{proposition}
\label{prop:nonexpansive}
For every $x=(q,v)\in\mathcal C$ and $z\in T_qQ$,
\begin{equation}
    \|\Pi_\Phi(x)z\|_{M(q)}
    \leq
    \|z\|_{M(q)}.
    \label{eq:nonexpansive}
\end{equation}
Consequently, if
$a^\star=c_\Phi+\Pi_\Phi g^\star$ and
$\widehat a_N=c_\Phi+\Pi_\Phi\widehat g_N$, then
\begin{equation}
    \|\widehat a_N(x)-a^\star(x)\|_{M(q)}
    \leq
    \|\widehat g_N(x)-g^\star(x)\|_{M(q)}.
    \label{eq:error_bound}
\end{equation}
\end{proposition}
\begin{proof}
Since $\Pi_\Phi$ is an $M(q)$-orthogonal projector,
$\Pi_\Phi^2=\Pi_\Phi$ and
$\Pi_\Phi^\top M=M\Pi_\Phi$. Decompose
$z
    =
    \Pi_\Phi z+(I-\Pi_\Phi)z$. The two terms are $M(q)$-orthogonal, since
$(\Pi_\Phi z)^\top
    M(q)(I-\Pi_\Phi)z=
    z^\top
    \Pi_\Phi^\top M(q)(I-\Pi_\Phi)z =
    z^\top
    M(q)\Pi_\Phi(I-\Pi_\Phi)z
    =0$. Therefore, by the Pythagorean identity,
$\|z\|_{M(q)}^2
    =
    \|\Pi_\Phi z\|_{M(q)}^2
    +
    \|(I-\Pi_\Phi)z\|_{M(q)}^2$, which proves~\eqref{eq:nonexpansive}.
For the estimator,
$\widehat a_N-a^\star
    =
    \Pi_\Phi
    (\widehat g_N-g^\star)$, since the same affine correction $c_\Phi$ appears in both fields.
Applying~\eqref{eq:nonexpansive} gives~\eqref{eq:error_bound}.
\end{proof}
For the posterior mean in Theorem~
\ref{thm:constraint_preservation}, let
$P_X:=\operatorname{diag}_i\Pi_\Phi(x_i)$ and
$C_X:=\operatorname{col}_i c_\Phi(x_i)$, and define
\begin{align}
\widetilde g_N(x)
:=
m_0(x)
+
K_0(x,X)\mathbf P_X^\top
\Big(
\mathbf P_XK_0(X,X)\mathbf P_X^\top+\Sigma
\Big)^{-1}
\bigl(Y-C_X-\mathbf P_Xm_0(X)\bigr).
\end{align}
Then
$\hat a_{\Phi,N}=c_\Phi+\Pi_\Phi\widetilde g_N$,
so Proposition~\ref{prop:nonexpansive} applies directly to the posterior mean of the
structure-preserving GP. Finally we show that the nonholonomic projection preserves the consistency of the ambient GP estimator.
\begin{theorem}
\label{thm:consistency}
Assume that $\mathcal X\subset\mathcal C$ is compact and that the induced
ambient estimator $\widetilde g_N$ is uniformly consistent in probability
for $g^\star$ in the kinetic-energy norm, i.e., for every $\varepsilon>0$,
\begin{align}
\mathbb P\!\left(
\sup_{x=(q,v)\in\mathcal X}
\|\widetilde g_N(x)-g^\star(x)\|_{M(q)}
>\varepsilon
\right)\rightarrow 0.
\end{align}
Then the posterior mean $\widehat a_{\Phi,N}$ is uniformly consistent in
probability for
$a^\star=c_\Phi+\Pi_\Phi g^\star$.
\end{theorem}
\begin{proof}
By Proposition~5 and the representation above, the following holds
$
\sup_{x\in\mathcal X}
\|\widehat a_{\Phi,N}(x)-a^\star(x)\|_{M(q)}
\le
\sup_{x\in\mathcal X}
\|\widetilde g_N(x)-g^\star(x)\|_{M(q)} .
$
Hence, for every $\varepsilon>0$,
\begin{align}
\mathbb P\!\left(
\sup_{x\in\mathcal X}
\|\widehat a_{\Phi,N}(x)-a^\star(x)\|_{M(q)}
>\varepsilon
\right)
\le
\mathbb P\!\left(
\sup_{x\in\mathcal X}
\|\widetilde g_N(x)-g^\star(x)\|_{M(q)}
>\varepsilon
\right),
\end{align}
and the right-hand side converges to zero by assumption.
\end{proof}
Since $\mathcal X$ is compact and $M(q)$ is continuous and positive
definite, there exist constants
$0<\underline m\leq\overline m<\infty$ such that
$\underline m I\preceq M(q)\preceq\overline m I$ on 
$\mathcal X$. Therefore,
\begin{align}
\sup_{x\in\mathcal X}
\|\widehat a_{\Phi,N}(x)-a^\star(x)\|_2
\leq
\sqrt{\frac{\overline m}{\underline m}}\,
\sup_{x\in\mathcal X}
\|\widetilde g_N(x)-g^\star(x)\|_2 .
\end{align}
Thus uniform consistency in the Euclidean norm also transfers from the
induced ambient estimator to the structure-preserving estimator.
Since $\Pi_\Phi$ and $c_\Phi$ are invariant under different defining maps
of the same admissible acceleration space, the error bounds and consistency
results above are representation independent.

\section{Three-Dimensional Flocking with Nonlinear Velocity Constraints}
\label{sec:flocking}

Velocity alignment is a central mechanism in collective-motion models,
from Reynolds' behavioral rules \citep{Reynolds1987} and the polar model
of~\cite{VicsekEtAl1995} to distributed flocking and
velocity-consensus models \citep{OlfatiSaber2006,CuckerSmale2007}.
Learning-based approaches have more recently recovered collective behavior
from data using reinforcement learning \citep{DurvePeruaniCelani2020},
neural ODEs \citep{JiahaoPanHsieh2022}, and Gaussian processes
\citep{BeckersPappasColombo2022}. Our objective is different: rather than
learning a feedback law that drives arbitrary velocities toward alignment,
we treat directional alignment itself as a nonlinear geometric constraint
and embed it directly into the learned dynamics. Cohesion, separation, and
other interaction mechanisms may remain part of the ambient dynamics, while
the alignment geometry is satisfied exactly.

\subsection{Flocking Geometry and Structure-Preserving GP on the Alignment Manifold}
\label{subsec:flocking_gp}

Consider $N$ agents with $q_i,v_i\in\mathbb R^3$ and
$Q=(\mathbb R^3)^N$, endowed with the kinetic-energy metric
$M=\operatorname{diag}_i(m_iI_3)$, $m_i>0$.
Let $\mathcal G=(\mathcal V,\mathcal E)$ be a connected undirected
interaction graph. Directional alignment between neighboring agents is
described by
$\Phi_{ij}(v):=v_i\times v_j=0$, $(i,j)\in\mathcal E$, and we collect
these equations into
$\Phi_{\mathcal G}(v):=
\operatorname{col}_{(i,j)\in\mathcal E}\Phi_{ij}(v)
\in\mathbb R^{3|\mathcal E|}$.
Unlike velocity consensus, this constraint requires only parallel
velocities and therefore allows the agents to have different speeds.

The set $\Phi_{\mathcal G}^{-1}(0)$ also contains zero and antiparallel
velocities. We consider the coherently aligned manifold
$\mathcal V_{\mathrm{al}}^{+}:=
\{v:\,v_i=\alpha_i d,\ d\in\mathbb S^2,\ \alpha_i>0,\ i=1,\ldots,N\}$
and set
$\mathcal C_{\mathrm{al}}:=Q\times\mathcal V_{\mathrm{al}}^{+}\subset TQ$.
For a connected graph, $\mathcal V_{\mathrm{al}}^{+}$ is the subset of the
constraint set where all velocities are nonzero and share the same
orientation. It has dimension $N+2$: the $N$ scalars $\alpha_i$ determine
the individual speeds, while the two degrees of freedom of
$d\in\mathbb S^2$ determine the common heading.

Let $A_{\mathcal G}(v):=D_v\Phi_{\mathcal G}(v)$. For an edge
$(i,j)\in\mathcal E$,
$D\Phi_{ij}(v)[\delta v]
=\delta v_i\times v_j+v_i\times\delta v_j$.
With $\widehat w z:=w\times z$, the corresponding nonzero blocks of the
Jacobian are $-\widehat v_j$ and $\widehat v_i$.
\begin{proposition}
\label{prop:flocking_rank}
Let $\mathcal G$ be connected and
$v\in\mathcal V_{\mathrm{al}}^{+}$, with $v_i=\alpha_i d$. Then
$\operatorname{rank}A_{\mathcal G}(v)=2(N-1)$ and
$T_v\mathcal V_{\mathrm{al}}^{+}
=
\ker A_{\mathcal G}(v)
=
\{
\delta v:\,
\delta v_i=\beta_i d+\alpha_i\xi,\;
\beta_i\in\mathbb R,\;
\xi\in d^\perp
\}$. Consequently, the $M$-orthogonal projector onto
$\ker A_{\mathcal G}(v)$ is the same for every connected interaction
graph on the same set of agents.
\end{proposition}
\begin{proof}
At $v_i=\alpha_i d$,
$D\Phi_{ij}(v)[\delta v]
=d\times(\alpha_i\delta v_j-\alpha_j\delta v_i)$.
Hence $D\Phi_{ij}(v)[\delta v]=0$ if and only if
$P_d^\perp\delta v_j/\alpha_j
=P_d^\perp\delta v_i/\alpha_i$, where
$P_d^\perp:=I_3-dd^\top$.
Since $\mathcal G$ is connected, these relations propagate along the
graph and there exists a common $\xi\in d^\perp$ such that
$P_d^\perp\delta v_i=\alpha_i\xi$ for every agent. The longitudinal
components remain arbitrary, so
$\delta v_i=\beta_i d+\alpha_i\xi$. Differentiating $v_i=\alpha_i d$, with
$\dot d\in d^\perp$, gives the same characterization of
$T_v\mathcal V_{\mathrm{al}}^{+}$. 
Thus $\dim\ker A_{\mathcal G}=N+2$ and
$\operatorname{rank}A_{\mathcal G}=3N-(N+2)=2(N-1)$.
Since this kernel depends only on the aligned state and not on the
particular connected graph, uniqueness of the $M$-orthogonal projector
gives the final statement.
\end{proof}
Proposition~\ref{prop:flocking_rank} also shows why the
pseudoinverse formulation of
Section~\ref{subsec:nonlinear_mechanics} is needed here.
The map $\Phi_{\mathcal G}$ contains $3|\mathcal E|$ scalar constraint
functions, while its Jacobian has rank only $2(N-1)$ on
$\mathcal V_{\mathrm{al}}^{+}$. Thus even a spanning-tree representation
contains dependent scalar equations because each three-dimensional
cross-product equation imposes only two independent conditions at an
aligned nonzero state. The Moore-Penrose formulation allows the natural
cross-product constraints to be retained without selecting local
components or pruning the interaction graph.

An additional consequence is that, on $\mathcal V_{\mathrm{al}}^{+}$, adding or
removing interaction edges does not alter the admissible infinitesimal
motion provided the graph remains connected. The graph determines how
alignment is represented through pairwise equations, whereas the resulting
tangent geometry depends only on the common aligned-velocity manifold.

Since $\Phi_{\mathcal G}$ depends only on velocity $v$, $b=0$ and
$c_{\Phi_{\mathcal G}}=0$. The dynamics and associated
matrix-valued kernel therefore reduce to
$a_{\mathcal G}(q,v)
    =
    \Pi_{\mathcal G}(v)g(q,v)$,
    $\Pi_{\mathcal G}
    =
    I-M^{-1}A_{\mathcal G}^\top
    (A_{\mathcal G}M^{-1}A_{\mathcal G}^\top)^\dagger
    A_{\mathcal G}$,
    $K_{\mathcal G}(x,x')
    =
    \Pi_{\mathcal G}(v)K_0(x,x')
    \Pi_{\mathcal G}(v')^\top$.

By Proposition~\ref{prop:flocking_rank}, both $\Pi_{\mathcal G}$ and
$K_{\mathcal G}$ are independent of the particular connected graph when
evaluated on $\mathcal V_{\mathrm{al}}^{+}$. Hence different connected
pairwise descriptions of the same aligned configuration generate the
same structured GP prior.

The tangent-space characterization also gives a direct mechanical
interpretation of the admissible accelerations. If $v_i=\alpha_i d$,
then any admissible acceleration has the form
$a_i=\beta_i d+\alpha_i\xi$, with $\xi\perp d$. The longitudinal
component $\beta_i d$ changes the speed of agent $i$ independently,
whereas the same transverse vector $\xi$ appears for every agent and
therefore produces a coordinated change of heading. Along an aligned
trajectory these variables are precisely $\beta_i=\dot\alpha_i$ and
$\xi=\dot d$.

For the diagonal metric above, this structure can be seen explicitly.
For $z\in\mathbb R^{3N}$,
$(\Pi_{\mathcal G}(v)z)_i
    =
    (d^\top z_i)d+\alpha_i\xi_z$ with 
$\displaystyle{\xi_z
    =
    \frac{\sum_{j=1}^N m_j\alpha_jP_d^\perp z_j}
         {\sum_{j=1}^N m_j\alpha_j^2}}$.
Thus the projection preserves each longitudinal acceleration while
replacing the transverse components by their unique mass-weighted
coordinated turning motion. In particular, the projection does not impose
equal speeds: it preserves exactly the degrees of freedom intrinsic to
directional flocking.

By Theorem~\ref{thm:constraint_preservation}, the constrained prior and
posterior predictions satisfy
$A_{\mathcal G}(v)a_{\mathcal G}(q,v)=0$ on
$\mathcal C_{\mathrm{al}}$. In particular, the posterior-mean second-order
vector field is tangent to $\mathcal C_{\mathrm{al}}$, so the
continuous-time learned dynamics preserve velocity alignment exactly.
The simulations below compare this structural preservation with standard
and projected GP baselines trained from the same data.

\subsection{Mechanical Benchmark and Robustness to Incompatible Observations}
\label{sec:flocking-sim}

We evaluate the proposed GP on a mechanical realization of the alignment
geometry above. Rather than prescribing the reference dynamics directly in
adapted coordinates, we consider potential-driven particles subject to the
nonlinear parallel-velocity constraint, in the spirit of the nonlinear
nonholonomic particle models of \cite{BlochRojo2016}.
For equal masses, $M=I$, we take
$L(q,v)=\frac12\sum_{i=1}^{N}\|v_i\|^2-V(q)$, with
$V(q)=\sum_{i<j}U(r_{ij})$, $r_{ij}=\|q_i-q_j\|$, and Morse interaction
$U(r)=C_r e^{-r/\ell_r}-C_a e^{-r/\ell_a}$, where
$C_r=2$, $C_a=1$, $\ell_r=0.45$, and $\ell_a=1.20$.
Thus $g^\star(q)=-\nabla V(q)$ provides short-range separation and
long-range cohesion, while alignment is not generated by an additional
steering term but by the nonlinear constraint
$\Phi_{ij}(v)=v_i\times v_j=0$.

The reference dynamics are therefore the Chetaev dynamics
$a^\star(q,v)=\Pi_{\mathcal G}(v)g^\star(q)$. For the sampled states,
$A_{\mathcal G}(v)g^\star(q)\neq0$ generically, so the reaction is active
and the benchmark is not obtained by choosing an already admissible ambient
field. For $N=4$, we additionally compute $a^\star$ from the redundant
Lagrange-d'Alembert-Chetaev equations and verify agreement with the
projector and null-space formulations to numerical precision.

The distinction between known geometry and unknown dynamics is important
here: $C_{\mathrm{al}}$, $\Phi_{\mathcal G}$, and hence $\Pi_{\mathcal G}$ are
assumed known, whereas the learner is not given $V$ or $g^\star$.
We use the translation-invariant input
$z=(\widetilde q,\alpha,d)$, with
$\widetilde q_i=q_i-N^{-1}\sum_j q_j$ and $v_i=\alpha_i d$, and the grouped
squared-exponential kernel
$k(z,z')=\sigma_f^2\exp[-\frac12(
\|\widetilde q-\widetilde q'\|^2/(N\ell_q^2)
+\|\alpha-\alpha'\|^2/(N\ell_\alpha^2)
+\|d-d'\|^2/\ell_d^2)]$.
We compare the unconstrained standard GP, the projected GP obtained by
conditioning the standard GP and projecting its posterior afterwards, and
the NH GP (GP with the proposed nonholonomic kernel of Proposition~\ref{prop:gp_pushforward}), which incorporates the projector before conditioning.
For the latter, we use Corollary~\ref{cor:adapted}, reducing the latent output dimension from
$3N$ to $N+2$.

To isolate the statistical effect of incorporating the geometry into the
prior, we perturb the same noisy acceleration observations by components
normal to the admissible space,
$y_i^{(\rho)}
=a^\star(x_i)+\varepsilon_i
+\rho_\perp\sqrt{3N}\,a_{\mathrm{rms}}\eta_i$,
$U(x_i)^\top\eta_i=0,\, \|\eta_i\|=1$,
where $U(x_i)$ is an orthonormal frame of
$\ker A_{\mathcal G}(v_i)$ and $a_{\mathrm{rms}}$ is the componentwise RMS
training acceleration. The isotropic term $\epsilon_i$ models ordinary measurement noise, while
the additional normal component is a controlled stress test that isolates
sensitivity to constraint-incompatible information.  Kernel and noise hyperparameters are calibrated once
from an independent uncontaminated data set and kept fixed throughout the
$\rho_\perp$ sweep, exactly matching the fixed-hyperparameter setting of the
normal-noise rejection result of Section~3.1.

\begin{figure}[b]
    \centering
    \vspace{-0.3cm}
    \includegraphics[width=\textwidth]
    {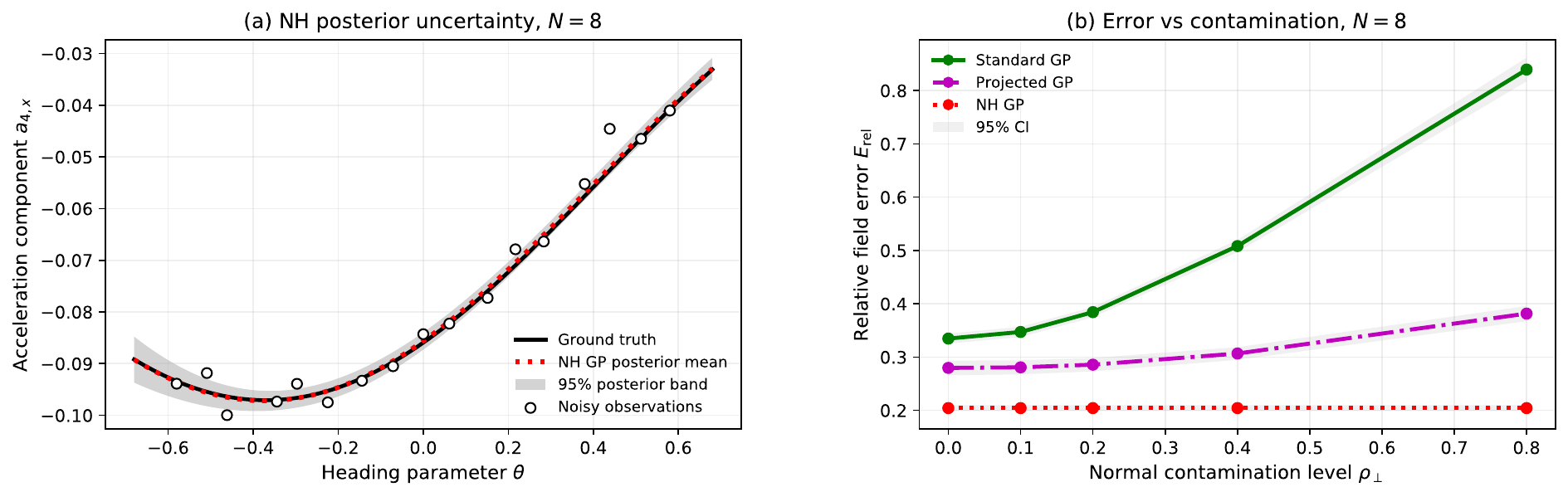}
\vspace{-0.7cm}\caption{
Nonlinear nonholonomic flocking benchmark. The proposed kernel (NH, red) outperforms the standard and projected GP.
(a) NH posterior mean with $95\%$ posterior band.
(b) Relative field error versus normal contamination.
}\vspace{-0.2cm}  \label{fig:flocking-benchmark1}
\end{figure}

\begin{figure}[t]
    \centering
    \includegraphics[width=\textwidth]
    {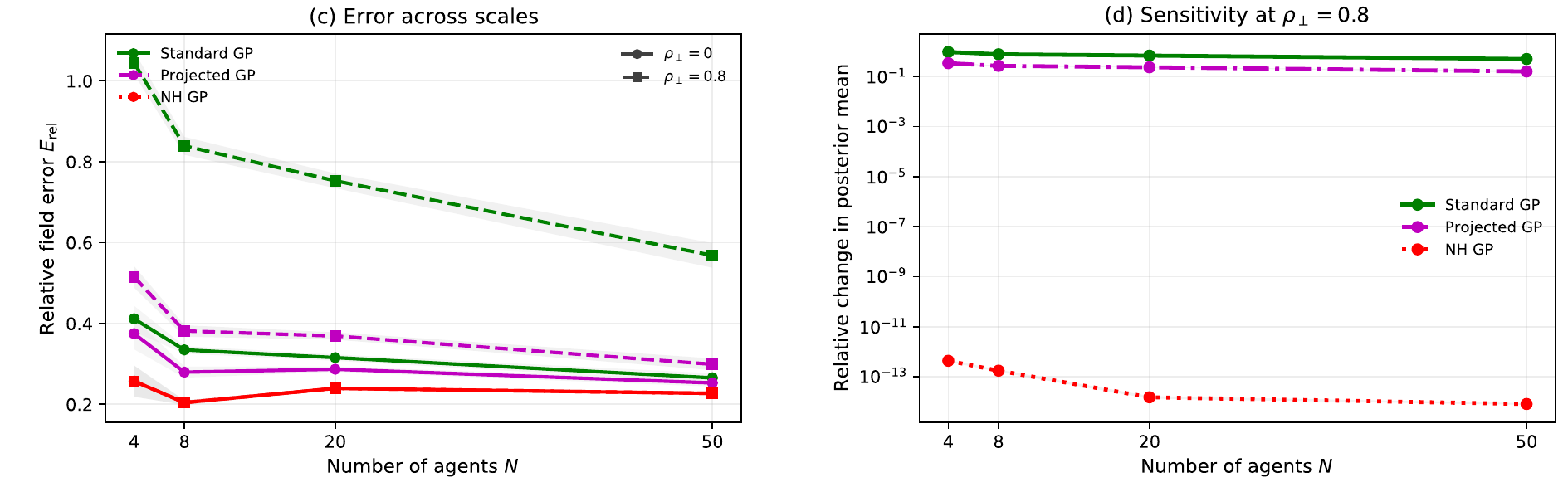}
\vspace{-0.9cm}\caption{
The proposed kernel (NH, red) demonstrates superior performance across varying numbers of agents.
(c) Relative field error across $N=4,8,20,50$ for
$\rho_\perp=0$ and $0.8$.
(d) Relative posterior-mean change at $\rho_\perp=0.8$.
Shaded bands are $95\%$ confidence intervals over five runs.
}\vspace{-0.3cm}  \label{fig:flocking-benchmark2}
\end{figure}

For stacked test-set accelerations $A^\star$ and predictions $\hat A_\rho$,
we use $E_{\mathrm{rel}}:=\|\hat A_\rho-A^\star\|_F/\|A^\star\|_F$ and
$\Delta_\rho:=\|\hat A_\rho-\hat A_0\|_F/\|A^\star\|_F$;
95\% confidence intervals are $\bar z\pm1.96\,s_z/\sqrt{5}$ over five
independent evaluation sets. Figures~\ref{fig:flocking-benchmark1} and~\ref{fig:flocking-benchmark2} summarize the numerical results.
In Figure~\ref{fig:flocking-benchmark1}, Panel~(a) shows a representative NH posterior, together with its
$95\%$ posterior band. Panel~(b) illustrates the effect of
constraint-incompatible observations for the intermediate case $N=8$:
normal contamination progressively degrades the standard and projected GPs,
while the NH prediction remains unchanged. At $\rho_\perp=0.8$, the
relative field errors are $0.839$, $0.382$, and $0.204$ for standard,
projected, and NH, respectively, while the relative change of the NH
posterior is only $1.8\times10^{-13}$. Thus, post-projection guarantees
admissibility of the final prediction but cannot remove incompatible
observation components that have already propagated through conditioning.

In Figure~\ref{fig:flocking-benchmark2}, Panels~(c)-(d) show that this mechanism persists as both the number of
agents and the redundancy of the constraint representation increase.
For the complete graph $K_N$, the natural cross-product description contains
$3N(N-1)/2$ scalar constraint equations, whereas on $C_{\mathrm{al}}$ its rank is
only $2(N-1)$ and the admissible dimension is $N+2$. Hence, for
$N=4,8,20,50$, the four representations contain respectively
$18,84,570,$ and $3675$ scalar equations, with ranks $6,14,38,$ and $98$.
Table~\ref{tab:flocking-results} reports the corresponding prediction errors.
Despite the increasing dimension and redundancy, the NH posterior changes by
only $10^{-13}$ to $10^{-15}$ under strong normal contamination, while both
baselines are affected. The larger-$N$ cases are structural scalability tests
rather than sample-complexity comparisons, since the number of training
states is reduced as $N$ increases.

\begin{table}[t]
\centering
\small
\setlength{\tabcolsep}{4pt}
\begin{tabular}{c c c c c}
\hline
$N$ & $N_{\mathrm{tr}}$ &
$E_{\mathrm{rel}}(0)$ &
$E_{\mathrm{rel}}(0.8)$ &
$\Delta_{\mathrm{NH}}(0.8)$ \\
\hline
4  & 160 & .412/.375/.257 & 1.045/.515/.257 & $4.4\times10^{-13}$ \\
8  & 120 & .335/.280/.204 & .839/.382/.204  & $1.8\times10^{-13}$ \\
20 & 80  & .316/.287/.240 & .753/.369/.240  & $1.5\times10^{-14}$ \\
50 & 60  & .265/.253/.227 & .568/.299/.227  & $8.3\times10^{-15}$ \\
\hline
\end{tabular}
\caption{\vspace{-0.1cm}
Prediction error and sensitivity to constraint-incompatible training-data
contamination. Entries in the two $E_{\mathrm{rel}}$ columns are ordered as
Standard/Projected/NH; values are means over five independent evaluation
sets.
}
\label{tab:flocking-results}
\vspace{-0.5cm}
\end{table}

\section{Conclusions and Future Work}
\label{sec:conclusions}

We introduced a structure-preserving GP framework for nonlinear
nonholonomic dynamics whose constraint functions need not be independent.
The Chetaev equations and Moore-Penrose pseudoinverse determine an affine
space of admissible accelerations, together with its kinetic-energy metric
projector and Chetaev correction, independently of the chosen constraint
representation. Embedding this geometry into the prior yields exact
constraint preservation before and after conditioning, an explicit RKHS
and adapted-coordinate characterization, and nonexpansive error transfer
in the kinetic-energy metric. Unlike post-projection, the constrained
prior rejects observation components normal to the admissible directions
before they propagate through conditioning. For three-dimensional
directional flocking, the cross-product constraints induce the same
tangent space and metric projector for every connected interaction graph,
and experiments up to $N=50$ demonstrate exact constraint preservation
and robustness to constraint-incompatible observations. Future work will
address learning away from the constraint manifold, time-varying
interaction topologies, and broader classes of nonlinear nonholonomic
systems.

\acks{L. Colombo acknowledges financial support from Grant PID2022-137909NB-C21 funded by MCIN/ AEI/ 10.13039/501100011033 and by iRoboCity2030-CM, Robótica Inteligente para Ciudades Sos\-tenibles (TEC-2024/TEC-62), funded by the Programas de Actividades I+D en Tecnologías en la Comunidad de Madrid. A. Bloch was partially supported by NSF grants  DMS-2103026 and DMS-2605296, and AFOSR grants FA
9550-22-1-0215 and FA 9550-23-1-0400.}

\bibliography{l4dc2026}

\end{document}